\documentclass[12pt]{article}

\newcommand{\be}{\begin{equation}}
\newcommand{\ee}{\end{equation}}
\newcommand{\bi}[1]{\vspace{-3mm} \bibitem{#1}}

\begin{document}

\begin{center}
{\it Modern Physics Letters B 21 (2007) 163-174}
\vskip 3 mm
{\large \bf The Fractional Chapman-Kolmogorov Equation}
\vskip 3 mm

{\large \bf Vasily E. Tarasov } \\

\vskip 3mm
{\it Skobeltsyn Institute of Nuclear Physics, 
Moscow State University, Moscow 119991, Russia}

E-mail: tarasov@theory.sinp.msu.ru
\end{center}

%%%\begin{abstract}
{\small
The Chapman-Kolmogorov equation with fractional integrals
is derived. An integral of fractional order 
is considered as an approximation of the integral on fractal. 
Fractional integrals can be used to describe the fractal media.
Using fractional integrals, the fractional generalization 
of the Chapman-Kolmogorov equation is obtained.  
From the fractional Chapman-Kolmogorov equation, the
Fokker-Planck equation is derived.}
%%%\end{abstract}

%%%\vskip 7mm
%%%PACS {05.45.Df; 47.53.+n; 05.40.-a }
%%%Keywords: Fokker-Planck equation, fractional integrals, 
%%%\vskip 5mm

%%%%%%%%%%%%%%%%%%%%%%%%%%%%%%%%%%%%%%%%%%%%%%%%%%%%%%%%%%%%%%%%%%%%%%%%%%
\section{Introduction}

Integrals and derivatives of the fractional order goes back 
to Leibniz, Liouville, Riemann, Grunwald, and Letnikov \cite{SKM}. 
Fractional analysis has found many
applications in recent studies in mechanics and physics.
The interest in fractional integrals and derivatives 
has been growing continually 
during the last few years because of numerous applications. 
In a fairly short period of time, 
the list of such applications has become long. 
It includes chaotic dynamics \cite{Zaslavsky1,Zaslavsky2},
mechanics of fractal and complex media 
\cite{Mainardi,Media,Physica2005},
physical kinetics \cite{Zaslavsky1,Zaslavsky7,SZ},
plasma physics \cite{CLZ,ZM,Plasmas},  
astrophysics \cite{CMDA},
long-range dissipation \cite{GM,TZ2}, 
non-Hamiltonian mechanics \cite{nonHam,FracHam},
and long-range interaction \cite{Lask,TZ3,KZT}.

The natural questions arises: What could be the physical
meaning of the fractional integration?
This physical meaning can be following:
the fractional integration can be considered as an
integration in some noninteger-dimensional space.
If we use the well-known formulas for dimensional 
regularizations \cite{Col}, then we get
that the fractional integration can be considered as an 
integration in the fractional dimension space \cite{nonHam}
up to the numerical factor
$\Gamma(\alpha/2) /[ 2 \pi^{\alpha/2} \Gamma(\alpha)]$.
This interpretation was suggested in Ref. \cite{nonHam}.
Fractional integrals can be considered as approximations 
of integrals on fractals \cite{Svozil,RLWQ}. 
In Ref. \cite{RLWQ}, authors proved that integrals
on a net of fractals can be approximated by fractional integrals.
Using fractional integrals, we derive the fractional generalization 
of Chapman-Kolmogorov equation \cite{Kolm1,Kolm2}. 
In this paper, the generalization of the Fokker-Planck equation 
for fractal media
is derived from the fractional Chapman-Kolmogorov equation. 

In Sec. 2, a brief review of 
the Hausdorff measure, the Hausdorff dimension 
and integration on fractals is carried out 
to fix notation and provide a convenient reference.
The connection of integration on fractals and fractional 
integration is discussed.
We derive the fractional generalization of the average values equation.
In Sec. 3, the fractional Chapman-Kolmogorov equation
is derived by using fractional integration. 
In Sec. 4, the fractional Fokker-Planck equation for the fractal media 
is derived from the suggested fractional Chapman-Kolmogorov equation. 
The stationary solutions of the Fokker-Planck equation
for fractal media are derived.

%%%%%%%%%%%%%%%%%%%%%%%%%%%%%%%%%%%%%%%%%%%%%%%%%%%%%%%%%%%%%%%%%%

\section{Integration on Fractal and Fractional Integration} 

Fractals are measurable metric sets 
with a non-integer Hausdorff dimension.  
Let us consider a brief review of 
the Hausdorff measure and the Hausdorff dimension  
in order to fix notation and provide a convenient reference. 

\subsection{Hausdorff measure and Hausdorff dimension}

Consider a measurable metric set $(W, \mu_H)$. 
The elements of $W$ are denoted by $x, y, z, . . . $, and represented by 
$n$-tuples of real numbers $x = (x_1,x_2,...,x_n)$ 
such that $W$ is embedded in $R^n$. 
The set $W$ is restricted by the conditions:
(1) $W$ is closed;
(2) $W$ is unbounded; 
(3) $W$ is regular (homogeneous, uniform) with its points randomly distributed.

The metric $d(x,y)$ as a function of two points $x$ and $y \in W$ 
can be defined  by
\be
d(x,y)=\sum^n_{i=1} |y_i-x_i| .
\ee
The diameter of a subset $E \subset W \subset R^n$ is 
\[
d(E)=diam(E)=sup\{ d(x,y): x , y  \in E \} .
\]

Let us consider a set $\{E_i\}$ of non-empty 
subsets $E_i$ such that
$dim(E_i) < \varepsilon$,  $\forall i$, and 
$W \subset \bigcup^{\infty}_{i=1} E_i$.
Then, we define 
\be
\xi(E_i,D)= \omega(D) [diam (E_i)]^D=\omega(D)  [d(E_i)]^D .
\ee
The factor $\omega(D)$ 
depends on the geometry of $E_i$, used for covering $W$.
If $\{E_i\}$ is the set of all (closed or open) balls in $W$, then
\be
\omega(D)=\frac{\pi^{D/2} 2^{-D}}{\Gamma(D/2+1)}. 
\ee

The Hausdorff dimension $D$ of a subset $E \subset W$  
is  defined \cite{Federer,R} by
\be \label{Hd}
D= dim_H (E)=\sup \{ d \in R: \  \mu_H (E, d ) = \infty  \} 
=\inf\{ d \in R : \ \mu_H(E,d)=0 \}.
\ee
From (\ref{Hd}), we obtain 
$\mu_H(E,d)=0$ for $d>D$; and
$\mu_H(E,d)=\infty$ for $d<D$.

The Hausdorff measure $\mu_H$ of a subset $E \subset W $ 
is \cite{Federer,R}:
\be
\mu_H(E,D)=
\lim_{\varepsilon \rightarrow 0} \inf_{\{E_i\}} 
\{ \sum^{\infty}_{i=1} \xi(E_i,D): 
\quad E \subset \bigcup_i E_i , \quad d(E_i)< \varepsilon \quad \forall i \} ,
\ee
or
\be 
\mu_H(E,D)= \omega(D) \lim_{d(E_i) \rightarrow 0} 
\inf_{\{E_i\}} \sum^{\infty}_{i=1} [d(E_i)]^D .
\ee
If $E \subset W$ and $\lambda >0$, then
$\mu_H(\lambda E,D) =\lambda^D \mu_H(E,D)$, 
where $\lambda E =\{ \lambda x, \ x \in E \}$.

%%%%%%%%%%%%%%%%%%%%%%%%%
\subsection{Function and integrals on fractal}

Let us consider the functions  
\be \label{f}
f(x)=\sum^{\infty}_{i=1} \beta_i \chi_{E_i}(x) , 
\ee
where $\chi_{E}$ is the characteristic function of $E$: 
$\chi_{E}(x)=1$ if $x \in E$, and $\chi_{E}(x)=0$ if $x \not \in E$.

The Lebesgue-Stieltjes integral for (\ref{f}) 
is defined by
\be \label{LSI}
\int_W f \, d \mu =\sum^{\infty}_{i=1} \beta_i \mu_H(E_i).
\ee
Therefore
\be \label{int}
\int_W f(x) \, d \mu_H (x) =
\lim_{ d(E_i) \rightarrow 0} \sum_{E_i} f(x_i) \xi(E_i,D)
=\omega(D) \lim_{ d(E_i) \rightarrow 0} \sum_{E_i} f(x_i) [d(E_i)]^D .
\ee
It is always possible to divide $R^n$ into parallelepipeds:
\be
E_{i_1...i_n} =\{  (x_1,...,x_n) \in W: x_j =
(i_j-1) \Delta x_j +\alpha_j, \
0 \le \alpha_j \le \Delta x_j, \quad j=1,...,n \} .
\ee
Then
\be
d \mu_H (x)= \lim_{d(E_{i_1...i_n}) \rightarrow 0}
\xi(E_{i_1 ... i_n},D)
=\lim_{d(E_{i_1 ... i_n}) \rightarrow 0}
\prod^n_{j=1} (\Delta x_j)^{D/n}=\prod^n_{j=1} d^{D/n} x_j .
\ee
The range of integration $W$ may also be parametrized by 
polar coordinates with $r=d(x, 0)$ and angle $\Omega$. 
Then $E_{r,\Omega}$ can be thought of as spherically 
symmetric covering around a center at the origin. 
In the limit, the function $\xi(E_{r,\Omega},D)$ gives
\be
d\mu_H(r,\Omega)=\lim_{d(E_{r, \Omega}) \rightarrow 0}
\xi (E_{r,\Omega},D)=d\Omega^{D-1} r^{D-1} dr. 
\ee

Let us consider $f(x)$ that is symmetric with respect 
to some point $x_0 \in W$, 
i.e. $f(x) = const$ for all $x$ such that $d(x, x_0)=r$ 
for arbitrary values of $r$. Then the transformation
\be \label{WrZ}
W \rightarrow W^{\prime} : \ x \rightarrow x^{\prime}=x-x_0
\ee
can be performed to shift the center of symmetry.  
Since $W$ is not a linear space, 
the transformation (\ref{WrZ}) need not be a map of $W$ onto 
itself, and (\ref{WrZ}) is measure preserving. 
Then the  integral over a $D$-dimensional metric space is 
\be \label{intWf}
\int_W f \, d\mu_H = \lambda(D) \int^{\infty}_0 f(r) r^{D-1} dr ,
\ee
where $\lambda(D)=2 \pi^{D/2} / \Gamma(D/2)$.
This integral is known in the theory of the 
fractional calculus \cite{SKM}. 
The right Riemann-Liouville 
fractional integral is
\be \label{FID}
(I^{D}_{-} f)(z)=\frac{1}{\Gamma(D)} \int^{\infty}_z (x-z)^{D-1} f(x) dx .
\ee
Then Eq. (\ref{intWf}) is reproduced by
\be \label{FIFI}
\int_W f \, d\mu_H = \frac{2 \pi^{D/2} \Gamma(D)}{\Gamma(D/2)} (I^{D}_{-} f)(0) .
\ee
Equation (\ref{FIFI}) connects the integral on fractal 
with integral of fractional order.
This result permits to apply different tools of the fractional calculus
\cite{SKM} for the fractal medium.
As a result, the fractional integral can be considered as an
integral on fractal 
up to the numerical factor $\Gamma(D/2) /[ 2 \pi^{D/2} \Gamma(D)]$.

Note that the interpretation of fractional integration
is connected with fractional dimension \cite{nonHam}.
This interpretation follows from the well-known formulas 
for dimensional regularization \cite{Col}.
The fractional integral can be considered as an
integral in the fractional dimension space
up to the numerical factor $\Gamma(D/2) /[ 2 \pi^{D/2} \Gamma(D)]$.
In Ref. \cite{Svozil}, it was proved that the fractal space-time 
approach is technically identical to the dimensional regularization.

\subsection{Properties of integrals}

The integral defined in Eq. (\ref{int}) satisfies the properties:

(1) Linearity:
\be \label{14}
\int_W (af_1+bf_2) \, d \mu_H=
a \int_W f_1 \, d \mu_H+b \int_W f_2 \, d \mu_H ,
\ee
where $f_1$ and $f_2$ are arbitrary functions; 
$a$ and $b$ are arbitrary constants.

(2) Translational invariance:
\be \label{15}
\int_W f(x+x_0) \, d \mu_H(x)= \int_W f(x) \, d \mu_H(x)
\ee
since $d \mu_H(x - x_0)=d \mu_H(x)$ as a consequence 
of homogeneity (uniformity).

(3) Scaling property:
\be \label{16}
\int_W f(\lambda x) \, d \mu_H(x)= \lambda^{-D}\int_W f(x) \, d \mu_H(x)
\ee
since $d \mu_H (x/\lambda)=\lambda^{-D}d \mu_H(x)$.

It is well-known \cite{Wilson,Col} that
conditions (\ref{14})-(\ref{16}) define the integral of 
the function $f(x) = \exp(-ax^2+bx)$ up to normalization:
\be \label{18}
\int_W \exp(-ax^2+bx) \, d \mu_H (x)=
\pi^{D/2} a^{-D/2} \, \exp(b^2/4a) .
\ee
For $b = 0$, Eq. (\ref{18}) is identical to the result 
that can be derived from (\ref{FIFI}) and 
is obtained directly without conditions (\ref{14})-(\ref{16}).

\subsection{Fractional average values}

The usual average value
\be <A>_1= \int^{+\infty}_{-\infty} A(x)\rho(x) dx  \ee
can be written as
\be \label{If}
<A>_1=\int^y_{-\infty} A(x)\rho(x) dx +\int^{\infty}_y A(x)\rho(x) dx . \ee
Using 
\be  \label{I+} (I^{\alpha}_{+}A)(y)=
\frac{1}{\Gamma (\alpha)} \int^{y}_{-\infty}
\frac{A(x)dx}{(y-x)^{1-\alpha}} , \ee
\be \label{I-} (I^{\alpha}_{-}A)(y)=
\frac{1}{\Gamma (\alpha)} \int^{\infty}_{y}
\frac{A(x)dx}{(x-y)^{1-\alpha}} , \ee
the average value (\ref{If}) can be present by
\be \label{A_1}
<A>_1=(I^{1}_{+}A\rho)(y)+(I^{1}_{-}A\rho)(y) . 
\ee
The fractional generalization of Eq. (\ref{A_1}) is 
\be \label{Aa} <A>_{\alpha}(y)=
(I^{\alpha}_{+}A\rho)(y)+(I^{\alpha}_{-}A\rho)(y) . \ee
Equation (\ref{Aa}) can be rewritten as
\be \label{FI2} <A>_{\alpha}(y)=
\int^{\infty}_{0} [ (A\rho)(y-x)+ (A\rho)(y+x)] d\mu_{\alpha}(x) , \ee
where 
\be \label{dm}
d\mu_{\alpha}(x)=\frac{|x|^{\alpha-1} dx}{\Gamma(\alpha)}=
\frac{d x^{\alpha}}{\alpha \Gamma(\alpha)} .\ee
Here, we use 
\be \label{xa}
x^{\alpha} =\beta(x) (x)^{\alpha}= sgn(x) |x|^{\alpha} , \ee
where $\beta(x)=(sgn(x))^{\alpha-1}$. 
The function $sgn(x)$ is equal to $+1$ for $x\ge0$,
and $-1$ for $x<0$.

To have the symmetric limits of the integral, 
we consider (\ref{FI2}) in the form
\be \label{FI5} <A>_{\alpha}(y)= \frac{1}{2}
\int^{+\infty}_{-\infty} [ (A\rho)(y-x)+ (A\rho)(y+x) ] d\mu_{\alpha}(x) . \ee
If $\alpha=1$, then we have the usual equation for the average value.

Let us introduce some notations to simplify Eq. (\ref{FI5}).
We define the integral operators
\be \label{hatI}
\hat {\it I}^{\alpha}_{x} f(x)=\frac{1}{2}
\int^{+\infty}_{-\infty} [ f(x)+ f(-x) ] d\mu_{\alpha}(x) .
\ee
Then (\ref{FI5}) has the form
\be \label{av1}
<A>_{\alpha}=\hat {\it I}^{\alpha}_{x} A(x)\rho(x) .
\ee
We will use the initial points that are set to zero ($y=0$).
Note that the fractional normalization condition is a special case
of this definition of average values: $<1>_{\alpha}=1$.

%%%%%%%%%%%%%%%%%%%%%%%%%%%%%%%%%%%%%%%%%%%%%%%%%%
\section{Fractional Chapman-Kolmogorov (FCK) Equation}

The Chapman-Kolmogorov equation \cite{Kolm1,Kolm2} 
may be interpreted as the condition of
consistency of distribution functions of different orders.
Kolmogorov \cite{Kolm1,Kolm2} derived a kinetic equation using a 
special scheme and conditions that are important for kinetics. 
Let $W(x,t;x_0,t_0)$ be a probability density of having a particle 
at the position $x$ at time $t$ if the particle was at the position $x_0$ 
at time $t_0\le t$. 

Denote by $\rho(x,t)$ the distribution functions for the given time $t$. 
Let us consider two well-known identities
\be  \label{rpr0} \rho(x,t) = \int^{+\infty}_{-\infty} d{x'}
\ W(x,t|x',t') \rho(x',t') , 
\quad \int^{+\infty}_{-\infty} \rho(x,t)=1 . \ee
Using the notation (\ref{hatI}), we can rewrite (\ref{rpr0})
in the form
\[  \rho(x,t) =\hat {\it I}^{1}_{x'}  \ W(x,t|x',t') \rho(x',t') , \quad
\hat {\it I}^{1}_x  \ \rho(x,t)=1 . \]
Then the fractional generalization of (\ref{rpr0}) is
\be  \label{rpr} \rho(x,t) =
\hat {\it I}^{\alpha}_{x'} \ W(x,t|x',t') \rho(x',t') . \ee
This equation is
the definition of conditional distribution function $W(x,t|x',t')$ 
referring to different time instants.
The normalization conditions for  the functions $W(x,t|x',t')$ 
and $\rho(x,t)$ are 
\be \label{nc-p} \hat {\it I}^{\alpha}_x  \ W(x,t|x',t')=1 , \quad
\hat {\it I}^{\alpha}_x  \ \rho(x,t)=1 . \ee
Substituting into the right-hand side of Eq. (\ref{rpr})
the value of $\rho(x',t')$
expressed via the distribution $\rho(x_0,t_0)$ at an earlier time,
\be  \label{rpr-2} \rho(x',t') =
\hat {\it I}^{\alpha}_{x_0}  \ W(x',t'|x_0,t_0) \rho(x_0,t_0) , \ee
we obtain the integral relation which includes the
intermediate point $x'$,
\be \label{20} \rho(x,t)=
\hat {\it I}^{\alpha}_{x'} \ \hat {\it I}^{\alpha}_{x_0}
 \  W(x,t|x',t')W(x',t'|x_0,t_0) \rho(x_0,t_0) . \ee
Using Eq. (\ref{20}), and Eq. (\ref{rpr}) in the form
\be  \label{rpr-3} \rho(x,t) =
\hat {\it I}^{\alpha}_{x_0}  \ W(x,t|x_0,t_0) \rho(x_0,t_0) , \ee
we derive a closed equation for transition
probabilities
\[ \hat {\it I}^{\alpha}_{x_0}  \ W(x,t|x_0,t_0) \rho(x_0,t_0) =
\hat {\it I}^{\alpha}_{x'}  \ \hat {\it I}^{\alpha}_{x_0}  \
W(x,t|x',t')W(x',t'|x_0,t_0) \rho(x_0,t_0) . \]
Since the equation holds for arbitrary $\rho(x_0,t_0)$,
we may equate the integrand. As the result, we obtain 
the fractional Chapman-Kolmogorov (FCK) equation 
\be W(x,t|x_0,t_0)=\hat {\it I}^{\alpha}_{x'}  \
W(x,t|x',t')W(x',t'|x_0,t_0) . \ee
This equation can be used to describes the Markov-type process 
in the fractal medium 
that is described by the continuous medium model \cite{Media}.

%%%%%%%%%%%%%%%%%%%%%%%%%%%%%%%%%%%%%%%%%%%%%%%%%%%%%%%%%%%%%%%%%%%%
\section{Fokker-Planck Equation from FCK Equation}

\subsection{Derivations of Fokker-Planck equation}

Let us consider the fractional average value (\ref{av1}).
Using Eq. (\ref{rpr}) in the form
\be  \label{rpr2} \rho(x,t) =
\hat {\it I}^{\alpha}_{x_0}  \ W(x,t|x_0,t_0) \rho(x_0,t_0) , \ee
we get 
\be \label{AAA}
<A>_{\alpha}=\hat {\it I}^{\alpha}_x  \ A(x)\ 
\hat {\it I}^{\alpha}_{x_0}  \ W(x,t|x_0,t_0) \rho(x_0,t_0) . \ee
We can rewrite Eq. (\ref{AAA}) as 
\be \label{av3} <A>_{\alpha}=\hat {\it I}^{\alpha}_{x_0}  \
\rho(x_0,t_0) \ \hat {\it I}^{\alpha}_x  \ A(x) \ W(x,t|x_0,t_0)  . \ee
We assume that $A=A(x^{\alpha})$, and use the Taylor expansion 
\be \label{TE2}
A(x^{\alpha})=A(x^{\alpha}_0+\Delta x^{\alpha})=A(x^{\alpha}_0)+
\left(\frac{\partial A(x^{\alpha})}{\partial x^{\alpha}} \right)_{x_0}
\Delta x^{\alpha}+ 
\frac{1}{2}\left(\frac{\partial^2 A(x^{\alpha})}{(\partial x^{\alpha})^2}
\right)_{x_0} (\Delta x^{\alpha})^2+..., \ee
where $x^{\alpha}=sgn(x)|x|^{\alpha}$ is defined by Eq. (\ref{xa}),
$\Delta x^{\alpha}=x^{\alpha}-x^{\alpha}_0$, and
\be \label{am}
\frac{\partial}{\partial x^{\alpha}}= \frac{|x|^{1-\alpha}}{\alpha}
\frac{\partial}{\partial x} .
\ee
If we use the usual Taylor expansion, then 
the integration by parts in Eq. (\ref{av3}) is more complicated. 
For the expansion (\ref{TE2}), the integration by parts 
in (\ref{av3}) can be realized in the simple form, 
\[ \hat {\it I}^{\alpha}_x B(x) 
\frac{\partial A(x^{\alpha})}{\partial x^{\alpha}}=
\int^{+\infty}_{-\infty} \frac{dx^{\alpha}}{\alpha \Gamma(\alpha)} B(x)
\frac{\partial A(x^{\alpha})}{\partial x^{\alpha}}=\]
\[ =\left( B(x)A(x) \right)^{+\infty}_{-\infty}-
\int^{+\infty}_{-\infty} 
\frac{dx^{\alpha}}{\alpha \Gamma(\alpha)} A(x^{\alpha})
\frac{\partial B(x)}{\partial x^{\alpha}} . \]
Substituting Eq. (\ref{TE2}) in Eq. (\ref{av3}), we get
\[ <A>_{\alpha}= \hat {\it I}^{\alpha}_{x_0}  \ A(x^{\alpha}_0) \rho(x_0,t_0)
\hat {\it I}^{\alpha}_x  \  W(x,t|x_0,t_0) +  \]
\[ +\hat {\it I}^{\alpha}_{x_0}  \
\left(\frac{\partial A(x^{\alpha})}{\partial x^{\alpha}}\right)_{x_0} 
\rho(x_0,t_0) \hat {\it I}^{\alpha}_x  \ \Delta x^{\alpha}  W(x,t|x_0,t_0)+\]
\be \label{av4}
+\frac{1}{2} \hat {\it I}^{\alpha}_{x_0}  \
\left(\frac{\partial^2 A(x^{\alpha})}{(\partial x^{\alpha})^2}\right)_{x_0}
\rho(x_0,t_0)  
\hat {\it I}^{\alpha}_x  \ (\Delta x^{\alpha})^2  W(x,t|x_0,t_0)+...  \ee
Let us introduce the functions:
\be \label{Pn} P_n(x_0,t,t_0)=\hat {\it I}^{\alpha}_x  \ (\Delta x^{\alpha})^n
W(x,t|x_0,t_0) . \ee
Using (\ref{Pn}) and (\ref{nc-p}), equation (\ref{av4}) gives 
\[ <A>_{\alpha}= \hat {\it I}^{\alpha}_{x_0} 
\ A(x^{\alpha}_0) \rho(x_0,t_0) +  
\hat {\it I}^{\alpha}_{x_0}  \
\left(\frac{\partial A(x^{\alpha})}{\partial x^{\alpha}}\right)_{x_0} 
\rho(x_0,t_0) P_1(x_0,t,t_0)+\]
\be \label{av5} +\frac{1}{2} \hat {\it I}^{\alpha}_{x_0}  \
\left(\frac{\partial^2 A(x^{\alpha})}{(\partial x^{\alpha})^2}\right)_{x_0}
\rho(x_0,t_0) P_2(x_0,t,t_0)+... \ .
\ee
Substitution of (\ref{av1}) in the form
\[ <A>_{\alpha}=
\hat {\it I}^{\alpha}_{x_0}  \ A(x^{\alpha}_0) \rho(x_0,t) , \]
into Eq. (\ref{av5}) gives 
\[ \hat {\it I}^{\alpha}_{x_0}  \
A(x_0)\left(\rho(x_0,t)-\rho(x_0,t_0)\right)=  \]
\be \label{av7} 
=\hat {\it I}^{\alpha}_{x_0}  \
\left(\frac{\partial A(x)}{\partial x^{\alpha}}\right)_{x_0} \rho(x_0,t_0)
P_1(x_0,t,t_0)+ \frac{1}{2} \hat {\it I}^{\alpha}_{x_0}  \
\left(\frac{\partial^2 A(x)}{(\partial x^{\alpha})^2}\right)_{x_0}
\rho(x_0,t_0) P_2(x_0,t,t_0)+... \ . \ee
Then we use so-called Kolmogorov condition \cite{Kolm1,Kolm2}, and 
assume that the following finite limits exist:
\[ \lim_{\Delta t \rightarrow 0} \frac{P_1(x,t,t_0)}{\Delta t}= a(x,t_0) , 
\quad
\lim_{\Delta t \rightarrow 0} \frac{P_2(x,t,t_0)}{\Delta t}= b(x,t_0) , 
\quad
\lim_{\Delta t \rightarrow 0} \frac{P_n(x,t,t_0)}{\Delta t}= 0 , 
\]
where $n=3,4,... $, and $\Delta t=t-t_0$.
It is due to the Kolmogorov conditions that irreversibility
appears at the final equation.
Multiplying both sides of Eq. (\ref{av7}) by $1/\Delta t$ and
consider the limit $\Delta t \rightarrow 0$, we obtain
\[ \hat {\it I}^{\alpha}_{x_0}  \ A(x^{\alpha}_0)
\left(\frac{\partial \rho(x_0,t)}{\partial t} \right)_{t_0} = \] 
\[ =\hat {\it I}^{\alpha}_{x_0}  \
\left(\frac{\partial A(x^{\alpha})}{\partial x^{\alpha}}\right)_{x_0} 
\rho(x_0,t_0) a(x_0,t_0) 
+\frac{1}{2} \hat {\it I}^{\alpha}_{x_0}  \
\left(\frac{\partial^2 A(x^{\alpha})}{(\partial x^{\alpha})^2}\right)_{x_0}
\rho(x_0,t_0) b(x_0,t_0) . \]
Integrating by parts, we obtain
\be \label{ip1}
\hat {\it I}^{\alpha}_{x}  \
\frac{\partial A(x^{\alpha})}{\partial x^{\alpha}} \rho(x,t) a(x,t) 
=- \hat {\it I}^{\alpha}_{x}  \ A(x^{\alpha})
\frac{\partial (\rho(x,t) a(x,t))}{\partial x^{\alpha}} , \ee

\be \label{ip2}
\hat {\it I}^{\alpha}_{x}  \
\frac{\partial^2 A(x^{\alpha})}{(\partial x^{\alpha})^2}
\rho(x,t) b(x,t)
=\hat {\it I}^{\alpha}_{x}  \ A(x^{\alpha})
\frac{ \partial^2 (\rho(x,t) b(x,t)) }{ (\partial x^{\alpha})^2 } . \ee
Here, we use
\[ \lim_{x\rightarrow \pm \infty} \rho(x,t)=0 .\]
Then 
\[ \hat {\it I}^{\alpha}_{x}  \ A(x^{\alpha})
\left( \frac{\partial \rho(x,t)}{\partial t}+
\frac{\partial (\rho(x,t)a(x))}{\partial x^{\alpha}}-\frac{1}{2}
\frac{\partial^2 (\rho(x,t)b (x)) }{ (\partial x^{\alpha})^2 }
\right)=0. \]
Since the function $A=A(x^{\alpha})$ is an arbitrary function, then 
we have
\be \label{FP} \frac{\partial \rho(x,t)}{\partial t}+
\frac{\partial (\rho(x,t)a(x,t))}{\partial x^{\alpha}} -\frac{1}{2}
\frac{\partial^2 (\rho(x,t)b (x,t))}{ (\partial x^{\alpha})^2 }=0  , \ee
that is the Fokker-Planck equation that corresponds to
the FCK equation. 
This equation is derived from the fractional generalizations of the
average value and fractional normalization condition, which use 
the fractional integrals.

\subsection{Stationary solutions}

For the stationary case, the Fokker-Planck equation (\ref{FP}) is
\be \label{sFP1} 
\frac{\partial (\rho(x,t)a(x,t))}{\partial x^{\alpha}} -\frac{1}{2}
\frac{\partial^2 (\rho(x,t)b (x,t))}{(\partial x^{\alpha})^2}=0  . \ee
This equation can be rewritten as
\be \label{sFP2} 
\frac{\partial}{\partial x^{\alpha}} \left( \rho(x,t)a(x,t)-\frac{1}{2}
\frac{\partial (\rho(x,t)b (x,t))}{\partial x^{\alpha}}\right)=0  . \ee
Then 
\be \label{sFP3} 
\rho(x,t)a(x,t) -\frac{1}{2}
\frac{\partial (\rho(x,t)b (x,t))}{\partial x^{\alpha}}=const  . \ee
Supposing that the constant is equal to zero, we get
\be \label{sFP4} 
\frac{\partial (\rho(x,t)b(x,t))}{\partial x^{\alpha}}=
\frac{2a(x,t)}{b(x,t)} (\rho(x,t)b(x,t)) , \ee
The solution of (\ref{sFP4}) is
\be \label{sFP6} 
\ln (\rho(x,t)b(x,t)) = \int \frac{2a(x,t)}{b(x,t)} dx^{\alpha} +const. \ee
As the result, we obtain
\be \label{sFP7} 
\rho(x,t)= \frac{N}{b(x,t)} \exp \ 2 \int \frac{a(x,t)}{b(x,t)} \, dx^{\alpha} , \ee
where the coefficient $N$ is defined by the normalization condition. \\

Let us consider the special cases of the solution (\ref{sFP7}). \\
(1) If $a(x)=k$ and $b(x)=-D$, then 
the Fokker-Planck equation (\ref{FP}) has the form
\be \label{FP-case1} \frac{\partial \rho(x,t)}{\partial t}+
k\frac{\partial \rho(x,t)}{\partial x^{\alpha}} +\frac{D}{2}
\frac{\partial^2 \rho(x,t)}{(\partial x^{\alpha})^2}=0 , \ee
and the stationary solution is
\be \label{case1} 
\rho(x,t)= N_1 
exp \left( - \frac{2 k |x|^{\alpha}}{D} \right) . \ee
\vskip 5 mm

\noindent
(2) If $a(x)=k|x|^{\beta}$ and $b(x)=-D$, then 
\be \label{FP-case2} 
\frac{\partial \rho(x,t)}{\partial t}+
k\frac{\partial  |x|^{\beta}\rho(x,t)}{\partial x^{\alpha}} +\frac{D}{2}
\frac{\partial^2 \rho(x,t)}{(\partial x^{\alpha})^2}=0  . \ee
The stationary solution is
\be \label{case2} 
\rho(x,t)= N_2 
exp \left( - \frac{2 \alpha k |x|^{\alpha+\beta}}{(\alpha+\beta)D} \right) . 
\ee
If $\alpha+\beta=2$, we have
\be \label{case2c} \rho(x,t)= 
N_2 exp \left( - \frac{\alpha k}{D} x^{2} \right) . \ee
\vskip 5 mm

\noindent
(3) If 
\[ a(x)=\frac{\partial U(x)}{\partial x^{\alpha}}=\frac{|x|^{1-\alpha}}{\alpha} 
\frac{\partial U(x)}{\partial x},  \] 
and $b=-D$, then 
\[ \rho(x,t)= N_4 exp \left( - \frac{U(x)}{D} \right). \]
\vskip 5 mm

Let us consider Eq. (\ref{FP})
with $a(x)=k|x|^{\alpha}$ and $b=-D$. 
The general solution  can be presented as
\[ \rho(x,t)=\sum^{+\infty}_{n=0}
\sqrt{\frac{k}{2^n n!\pi D}} e^{-k x^{2\alpha}/ D} 
H_n(x^{\alpha \sqrt{k/ D}}) e^{-nkt} A_n , \]
where
\[ A_n=\sqrt{\frac{1}{2^n n!}}\  \hat {\it I}^{\alpha}_x \ p(x,0) 
H_n(x^{\alpha}\sqrt{k/ D}) . \]
The stationary solution is
\[\rho(x)=\left(\frac{k}{\pi D}\right)^{1/2} e^{-k x^{2\alpha}/ D} . \]

%%%\newpage
%%%%%%%%%%%%%%%%%%%%%%%%%%%%%%%%%%%%%%%%%%%%%%%%%%%%%%%%%%%%%%%%%%%%%%%
\section{Conclusion}

The concept of fractional integration
provides an approach to describe the fractal media. 
The fractional integrals can be used in order to formulate  
the dynamical equations in the fractal media. 
The fractional integration approach is potentially more useful 
for the physics of fractal media than traditional methods that 
use the integer integration.
Using fractional integrals, 
we derive the fractional generalization of the 
Chapman-Kolmogorov equation.
This equation can be used to describes the Markov-type process 
in the fractal medium 
that is described by the continuous medium model \cite{Media}. 
The fractional Chapman-Kolmogorov equation can have 
a wide application since it uses 
a relatively small number of parameters that can define 
a fractal medium of great complexity and rich structure.
In this paper, we derive the Fokker-Planck equations 
for the fractal media from the suggested 
fractional Chapman-Kolmogorov equation.

%%%%%%%%%%%%%%%%%%%%%%%%%%%%%%%%%%%%%%%%%%%%%%

%%%%%%%%%%%%%%%%%%%%%%%%%%%%%%%%%%%%%%%%%%%%%%%%%%%%%%%%%%%%%%%%%%%%%%%%%%

\end{document}